# Human Emotion Recognition Based On Galvanic Skin Response signal Feature Selection and SVM

Di Fan, Mingyang Liu, Xiaohan Zhang, Xiaopeng Gong
Jilin University Instrument Science and Engineering Institute,Changchun,130021,China

*Abstract*—A novel human emotion recognition method based on automatically selected Galvanic Skin Response (GSR) signal features and SVM is proposed in this paper. GSR signals were acquired by e-Health Sensor Platform V2.0. Then, the data is de-noised by wavelet function and normalized to get rid of the individual difference. 30 features are extracted from the normalized data, however, directly using of these features will lead to a low recognition rate. In order to gain the optimized features, a covariance based feature selection is employed in our method. Finally, a SVM with input of the optimized features is utilized to achieve the human emotion recognition. The experimental results indicate that the proposed method leads to good human emotion recognition, and the recognition accuracy is more than 66.67%.

## I. INTRODUCTION

GSR signal can be regarded as an effective tool applied for the emotional recognition which is a significant role in affective computing research field. It changes with the change of the body's sympathetic nervous system, depending on the body's secreted response of sweet glands [1]. The Galvanic Skin Response (GSR) signal, as an instruction of the skin conductivity, comes into being at the situation when the changes of the skin resistance owing to changes of contraction and dilation of blood vessels in the skin and the sweat gland secretion. Related researchers generally agreed that the changes of GSR can trigger the level of emotional arousal.

The researchers usually used the statistical characteristics of the GSR physiclogical signal within a period of time for feature extraction and emotion recognition. And we refer the method of feature extraction proposed by Universitat Augsburg [2], we extract the best representative statistic value of changes with GSR signal as the primitive character for researching emotional recognition in time domain and frequency domain. In order to judge the emotional status well from the GSR signal, we use the covariance statistical analysis to sift the characteristic variables from the 30 features. These 30 features has the problem that some features' correlation is high and it leads the recognition rate is low. In this way, a method of covariance is put forward to solve this problem.

In some realistic problems, some factors can not be controlled by humans, however, their different levels can surely have significant effects on measurable variables. The variance analysis has the disadvantage that if we only concentrate on the other factors' effects without observing these factors, the results will be inaccurate with the inappropriate weight of factors. Thus, we need to get rid of the other factors' influence on the analysis conclusion so as to study the influence of controlled variables in different level accurately. The analysis of covariance regard the factors that can not be controlled by humans as covariant, and at the situation of eliminating covariants' effects on observed situation, analyze the controlled variables' effects on observed situation, in this way we can judge the controlled variable precisely. Briefly, the variance analysis combines linear regression and variance analysis, and examine whether two or more groups correction factors have difference or not. This method can improve statistical power by eliminating the confounder's influence on analysis index [3].

## II. DATA ACQUISITION

To recognize emotion based on GSR signal, we collect the GSR signal in a particular emotional state first. In our study, we used short segments of movies and short films to elicit the target emotions, and used e-Health Sensor Platform V2.0 and MATLAB for Arduino to record the emotional physiological signals to set up the database of our experiment.

All subjects are Junior in Jilin University of Changchun, China. Materials for eliciting emotions are 4 video fragments cut elaborately from large amounts of movies representing 4 different emotions, such as happiness, grief, anger, and fear. Each fragment last 4-5 minutes and was connected with a short video which also last 4-5 minutes for calm recovery. And we add the emotion when people stay in calm. So there are 5 emotions that make the pattern recognition.

## III. FEATURE EXTRACTION

A. Pre-Processing

The raw GSR signals collected from experiments were firstly smoothed using wavelet denoising. The spectrum of GSR signal which won't overlap with that of other interference signals concentrates in the band 0.08-0.2Hz [4]. The Wavelet Function is just like a band pass filter, discrete wavelet transform uses the low pass filtering and high pass filtering to decompose the signal to different scales which is the frequency band of the signal. In our experiment, the signals are decomposed into five multiresolution levels and we use the db5 wavelet function. In order to remove individual difference, the data collected under emotion calm is used to normalize GSR signal under other emotions.

The normalized formula is defined as:

$$X_i^{'} = \frac{X_i - X_{i\min}}{X_{i\max} - X_{i\min}} \qquad (1)$$

Where $X_i$ is the i-th GSR initial characteristic data, $X_{imin}$ is the minimum value of this dimension characteristic data, $X_{imax}$ is the max value of this dimension characteristic data, $X_i'$ is the characteristic value after normalized.

B. Feature Extraction

Referring to 6 time-domain features mentioned in [5]; we extract 30 statistical features, 24 of which are time domain and the rest frequency one. The statistical features added mainly include range, maximum and minimum which can also reflect the change of GSR. It is convinced that the analysis of covariance can be applied to clarify the correlation between each feature and the others. The covariance difference value is smaller, the correlation between the features is smaller, the effect of this feature applied for recognition is better.

The covariance and correlation coefficient are defined as:

$$Cov(X,Y) = E(X-EX)(Y-EY) \quad (2)$$

$$\rho_{XY} = \frac{Cov(X,Y)}{\sqrt{DX} \cdot \sqrt{DY}} \quad (3)$$

The covariance $Cov(X,Y)$ is a quantity to describe the correlation between random variable $X$ and $Y$. For any $X$ and $Y$ two random variables, we have the conclusion that the amount of $Cov(X,Y)$ is bigger, the correlation between $X$ and $Y$ is bigger. The correlation coefficient $\rho_{XY}$ is a quantity to describe the correlation between random variable $X$ and $Y$, and it is a dimensionless quantity which means it can not be affected by the unit when it describes the linear relationship between $X$ and $Y$.

When we use the 30 features to make the pattern recognition about the emotion from GSR signals, the covariance can do a favor of picking up the features of low correlation which means the features are not similar. Under the circumstance of different emotion pattern recognition, the low correlation can make the recognition rate larger.

From Fig.1 and Fig.2, we can find the feature No.3,5,7,8,9,10,11,15,16,17,18,19,23,24 and 28 have the relatively low correlation. So, we use these 15 features to make the pattern recognition.

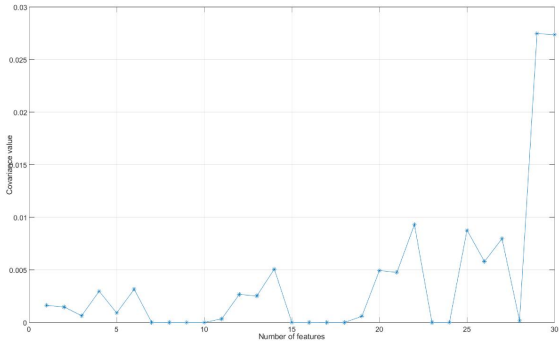

Figure 1. The covariance between the 30 dimensional features.

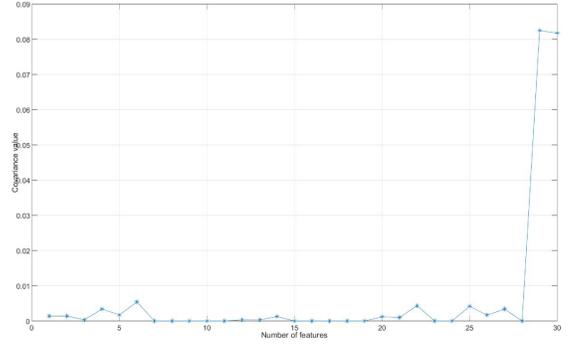

Figure 2. The covariance within the different emotion between 30 dimensional features of each feature.

Support Vector Machine (SVM) is a set of machine learning approaches used for classification and regression. This is a consequence of the optimization algorithms used for parameter selection and the statistical measures used to select the best model. SVM is based on the concept of decision planes that define decision boundaries. A decision plane is one that separates between a set of objects having different class memberships [6].

To forecast the emotion of the input GSR signals, a support vector machine is utilized after the features gained. The value of emotion has 5 levels (happiness, grief, fear, anger, and calm) Thus, the selected SVM is a multi-class classification. Given the training vectors $x_i \in R^n, i=1,2,...l$, the C-support vector classification (C-SVC) is used to solve the optimization problem. In the $k$-class problem, let $(x_i,y_i)$ be a samples of unknown function, $x_i$ is the input (the features in our problem) of the function, $y_i \in \{1,...k\}$ is the class (the emotion value in our problem) of the samples, the optimization problem is to minimize:

$$\phi(w,\xi) = \frac{1}{2}\sum_{m=1}^{k}(w_m \cdot w_m) + C\sum_{i=1}^{l}\sum_{m \neq y_i}\xi_i^m \quad (4)$$

with constraints:

$$(w_{y_i} \cdot x_i) + b_{y_i} \geq (w_m \cdot x_i) + b_m + 2 - \xi_i^m$$
$$m \in \{1,...k\}, \xi_i^m \geq 0, i=1,...l \quad (5)$$

$C$ is a weight coefficient, $w,b$ are the coefficients of line equation. Then the decision function given in (after finding the saddle point of the Lagrangian) is:

$$L(x,a) = \arg\max_m [\sum_{i:y_i=m} A_i K(x_i,x_j) - \sum_{i:y_i \neq m} a_i^m K(x_i,x_j) + b_m]$$

$$0 \leq a_i^m \leq C, A_i = \sum_{m=1}^{k} a_1^m$$

(6)

where $K(x_i,y_i)$ is the kernel function of the SVM, there are four basic kernels: linear, polynomial, radial basis function

(RBF) and sigmoid. We used the polynomial as the kernel function:

$$K(x_i, x_j) = (\eta x_i^T x_j + r), \eta > 0 \quad (7)$$

where *r* is a initialization coefficient. In experiments, the effect of different kernels on the forecasting accuracy was analyzed and the cross validation was also employed. Fig.3 is the flow diagram of the proposed method to recognize the emotion.

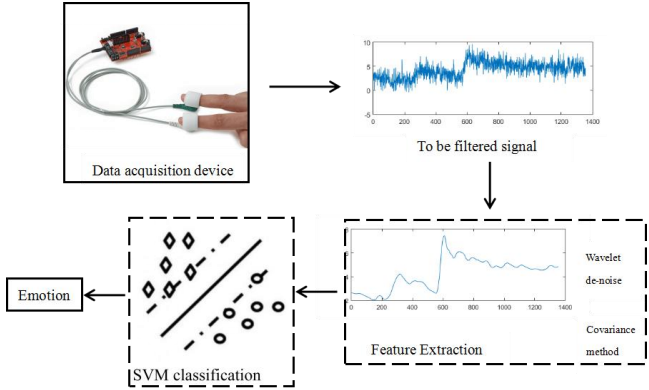

Figure 3. Block diagram of the proposed method.

The proposed algorithm consists of two parts, one is the training part, and another is the forecasting part. The training dataset is used to gain the SVM, then the trained SVM can be utilized in the testing part to achieve the Emotion forecasting. The detailed process is summarized as:

1) Collect GSR signal with the sensor and e-Health Sensor Platform V2.0;
2) Use wavelet function db5 to de-noise the GSR signal and normalize by Eq. (1);
3) Extract 30 dimension features from the processed GSR signal;
4) Select features by using covariance method by Eq. (2);
5) Train the SVM or make emotion recognition (after finishing the SVM training part).

## IV. EXPERIMENTS AND RESULTS

A. Preparation for Experiments

Before attending experiment, subjects should first fill in their true information, sign a piece of approval letter which represents their voluntary participation, and then wash their hands before GSR electrodes being attached to their forefingers and middle fingers. [7]

B. Results

Emotion database contains 257 emotion files for five emotion classes. Emotion classes happiness, grief, fear, anger, and calm are having 57, 51, 47, 43 and 59 GSR signals respectively. The LIBSVM[8] is trained on 15 feature vectors which using RBF kernel function. The Confusion matrices using RBF kernel for GSR pattern recognition accuracy with different number of features with different data are shown in Table I and II [9].

Table 1 CONFUSION MATRIX OF RBF CLASSIFIER FOR GSR PATTERN RECOGNITION

| Target Emotion | one to more Emotion Recognition(%) | |
|---|---|---|
| | Number of feature | |
| | 15 | 30 |
| Happiness | 33.33 | 66.67 |
| Grief | 66.67 | 33.33 |
| Fear | 100 | 33.33 |
| Anger | 66.67 | 66.67 |
| Calm | 66.67 | 33.33 |

Table 2 CONFUSION MATRIX OF RBF CLASSIFIER FOR GSR PATTERN RECOGNITION ACCURACY WITH DIFFERENT NUMBER OF FEATURES WITH DIFFERENT DATA

| Data | Accuracy(%) | |
|---|---|---|
| | Number of features | |
| | 15 | 30 |
| Training data | 73.57 | 83.57 |
| Test data | 66.67 | 46.67 |

From the Table 1 and 2, we can conclude that the method of sifting features with covariance works and get higher accuracy. To improve the accuracy of the experiment, then we use the cross validation to test the result. The purpose of using cross validation is to get a more stable and reliable model, the main theory of cross validation is to divide the dataset into train set and validation set. First, use the train set to the train classifier. Then, use the validation set to test the model and judge the classifier.

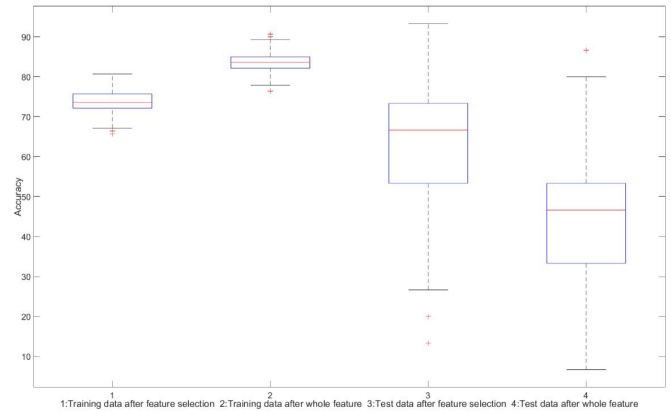

Figure 4. Results of the cross validation

From Fig.4, we can suppose that the model we build up works well and it validates the result we get from the LIBSVM.

## V. CONCLUSIONS

In this paper, emotion database of GSR signal is used for feature extraction. 15 features are selected with the covariance method. From experiment results, it is observed that the emotion can be detected from GSR signal and results from LIBSVM by using RBF kernel function is 66.67%. Covariance is a good way to help to make the emotion recognition and there will be more efficient with some new features added to help do the recognition.

ACKNOWLEDGEMENTS

This work is supported by Jilin University's University Student Renovation Project (2015650950).